\documentclass[manuscript]{article}

\usepackage{amsmath}
\usepackage{amsfonts}
\usepackage{amsthm}
\usepackage[ruled,vlined,commentsnumbered]{algorithm2e}
\usepackage{algorithmic}

\usepackage{mathtools}
\usepackage[square,sort,comma,numbers]{natbib}
\usepackage{hyperref}
\usepackage{color}

\usepackage[legalpaper, margin=1.5in]{geometry}

\newtheorem{theorem}{Theorem}
\newtheorem{corollary}[theorem]{Corollary}

\newtheorem{lemma}[theorem]{Lemma}

\newcommand{\congest}{CONGEST\xspace}

\newenvironment{apthm}[1]{\par\addvspace{3mm}\noindent\textbf {Theorem~\ref{#1}\;}}{\par\addvspace{3mm}}

\title{\textbf{Deterministic \congest Algorithm for MDS on Bounded Arboricity Graphs}}
\author{Saeed Akhoondian Amiri\thanks{University of Cologne, Germany, amiri@cs.uni-koeln.de.}}

\begin{document}
	\maketitle
		
	\begin{abstract}
		We provide a deterministic \congest algorithm to constant factor approximate the minimum dominating set on graphs of bounded arboricity in $O(\log n)$ rounds. 
		This improves over the well known randomized algorithm of Lenzen and Wattenhofer~\cite{ds-arbor} by making it a deterministic algorithm. 
	\end{abstract}

	\section{Introduction}
	The dominating set problem is to find a set of vertices whose closed neighborhood is the entire graph. Naturally, we are interested in a set of minimum size, or a so called minimum dominating set (MDS). The problem has been deeply studied in the sequential model as well as distributed models. 
	
	In this work, we are interested in algorithm design for LOCAL and \congest models. In these models every node of the network can be seen as a processor that communicates with its neighbors in synchronous communication rounds via communication links (the edges of the graph). In the LOCAL model there is no bandwidth limitation for communication links, but in the \congest model every link can carry a message of size $O(\log n)$ per round, where $n$ is the number of processors in the network. The aim of a distributed algorithm in such models is to provide an approximate solution for the given problem while minimizing the number of communication rounds. At the end every node outputs its share of the solution. For instance for the dominating set problem, every node outputs whether it belongs to the computed dominating set.
	
	From the distributed perspective, Kuhn et al.~\cite{Kuhn:2016} showed that in general graphs it is impossible to find a logarithmic approximate of  MDS in roughly speaking sublogarithmic rounds. From the positive side, Deurer et al.~\cite{DeurerKM19} provided a deterministic \congest algorithm with $O(\log \Delta)$-approximation guarantee in polylogarithmic number of rounds (by employing the recent breakthrough result for network decompositions~\cite{RozhonG20}).
	
	The status of approximation algorithms for the MDS problem in sparse graphs is much brighter. In particular, Lenzen et al.~\cite{ds-arbor}, provided a deterministic constant factor approximation LOCAL algorithm for planar graphs in a constant number of rounds. Wawrzyniak~\cite{wawrzyniak2013brief} extended that result to the \congest model and improved the approximation guarantee. The earlier work on planar graphs have been extended to bounded genus graphs by Amiri et al~\cite{amirilog,Amiri2016,saeed}. Then Czygrinow et al., first improved that algorithm to the \congest model, then they extended it to excluded minor graphs~\cite{CzygrinowHWW19,excludedminor}. Amiri et al.~\cite{amiri2018}, additionally provided a constant factor approximation on bounded expansion graphs in logarithmic rounds for a generalized version of the problem: the distance-$r$ MDS. The latter recently has been improved in two directions: Kublenz et al.~\cite{sebi}, reduced the number of rounds to a constant but only for the standard MDS-problem, Amiri and Wiederhake~\cite{ciac} showed that for high girth graphs the approximation algorithm for distance $r$-MDS can be obtained by $O(r)$ rounds. The minimum girth requirement showed to be useful in a recent work of Alipour and Jafari~\cite{AlipourJ20}, where they showed that in $C_4$-free planar graphs there is a better approximation guarantee for the MDS problem. 
	
	In another work, Lenzen and Wattenhofer provided a constant factor approximation randomized algorithm with a logarithmic number of rounds on graphs of bounded arboricity~\cite{ds-arbor}.
	This is the most generic result on sparse graphs, in a sense that it is a superclass of all aforementioned sparse graphs. Although the number of rounds is not constant and the algorithm is not deterministic, it performs much better than the existing work for general graphs. They also provided a deterministic  $O(\alpha\log \Delta)$-approximation algorithm in $O(\log \Delta)$ rounds, where $\alpha$ is the arboricity of the input graph and $\Delta$ is the maximum degree, they also show that their algorithm can fall in a worst case with approximation factor $O(\log \Delta)$. In this work we will extend their first result and show that it actually can be turned into a deterministic algorithm.
	
	\subsubsection*{Bottlenecks of Designing Algorithms for Bounded Arboricity Graphs}
	The edge set of a graph of bounded arboricity can be decomposed into a bounded number of forests. Such a decomposition is called \emph{Forest Decomposition} and it can be computed in logarithmic number of rounds. This is where the logarithmic barrier appears in designing algorithms on such graphs. 
	
	In the following we briefly explain why we need such a, relatively slow to compute, decomposition and why we cannot simply employ the known techniques for other sparse graph classes, at least for the case of MDS problem.
	
	All of the existing analysis methods for approximate MDS on planar, bounded genus, excluded minor, and bounded expansion graphs, are basically contraction-based: e.g.\ after contraction of subgraphs of a planar graph the resulting graph is still planar. In other words, all of the above-mentioned classes are closed under contraction. This facilitates analyzing domination problems: to our best of understanding in all of such existing analysis, a crucial point of the analysis is to contract certain subgraphs to a single vertex and since the new graph has the same structural attributes as the original one (e.g.\ it is still planar), we can employ the structural properties of them in the rest of analysis, for instance we can argue the resulting graph has a linear number of edges w.r.t.\ remaining vertices. 
	
	Thus, contraction based arguments  enabled researchers to employ the locality at its best. Such a delightful behavior does not appear in graphs of bounded arboricity. One famous example for this is the subdivided clique. Subdivided clique is a graph obtained by first taking a complete graph, then placing one vertex in the middle of each edge, or in other words, subdivide every edge once. Subdivided clique has arboricity $2$ but contracting half of the edges, makes the remaining a complete graph.
	
	Hence, the logarithmic barrier for graphs of bounded arboricity seems to be inevitable by using existing techniques. There is a progress in providing a decomposition with as fewest forests as possible~\cite{su1,GhaffariS17}. Breaking the logarithmic round complexity barrier for a constant number of forests contradicts the known lower bound of Linial~\cite{Linial87} for coloring unoriented trees. Working on a non-constant number of forests, happens to be useful in designing distributed coloring algorithms, however for approximation algorithms, in particular the case of the dominating set problem, it blows up the approximation guarantee.
	
	\subsubsection*{Our Contribution and the Algorithm of Lenzen and Wattenhofer}
	Our algorithms is the continuation of the idea that appeared first in~\cite{ds-arbor}. Lenzen and Wattenhofer did the following: first, they have computed the forest decomposition of the input graph. Based on this decomposition they create a specific auxiliary graph and conclude that the problem boils down to computing a maximal independent set (MIS) on that graph. Then by employing the randomized algorithm of Luby~\cite{Luby86} for MIS, they solved the problem in $O(\log n)$ rounds. Interestingly in their work, they started to use arguments based on set cover, however, due to the lack of existing distributed algorithms for set cover at the time they employed MIS\footnote{However, in~\cite{EvenGM18} authors mentioned that computing an approximate set cover in an instance with bounded frequency in logarithmic rounds is simple and they believe that it was known by the community prior to their paper, thus their focus was on breaking the logarithmic barrier.}.
	
	In this work, we employ the distributed set cover directly and consequently provide a constant factor approximation for MDS in $O(\log n)$ deterministic rounds of the \congest model. We also provide a simpler analysis than the previous work. Their analysis was based on separate counting arguments for child and parent nodes of MDS. We carry them out all together, which makes the proof of our following main theorem simpler. At the end we break the logarithmic barrier of the number of rounds by sacrificing the approximation guarantee.
	
	\noindent\begin{theorem}\label{thm:algorithm}
		There is a \congest algorithm that runs in $O(\log n)$ rounds and computes an $O(\alpha^2)$-approximation of MDS in graphs of arboricity at most $\alpha$.
	\end{theorem}

	\section{Useful Notation and Tools}
	We assume familiarity of reader with basic graph notations. First, we briefly explain the concept of forest decompositions.
	
	A graph $G$ has arboricity at most $\alpha$ if there are at most $\alpha$ spanning forests (or similarly spanning trees) such that their union spans all the edges of $G$. The set of such forests $\mathcal{F}$ is called a \emph{forest decomposition} of $G$. Barenboim and Elkin~\cite{BarenboimE10} provided a \congest algorithm that in $O(\log n/\epsilon)$ rounds computes a forest decomposition with at most $(2+\epsilon)\alpha$-forests in graphs of arboricity $\alpha$ (Algorithm 2 in the mentioned paper). The algorithm not only computes the decomposition but also calculates the directed rooted trees of each forest, hence by the end of the algorithm, every node knows to which forests it belongs. Additionally, every node knows its children and parents in the forests.
	
	The next ingredient we employ is a distributed algorithm for the distributed set cover problem with bounded frequency. In a set cover problem, there is a universe $U$ and a set $\mathcal{S}$ of subsets of $U$. The question is to find a subset of $\mathcal{S}$ such that the union of its elements covers $U$. Clearly, we would like to minimize the size of such a subset. In the \emph{bounded frequency} variant of the problem, every element of $U$ belongs to at most $f$ elements of $\mathcal{S}$, for some constant $f$.
	
	In the distributed variant we model the problem with a bipartite graph $H=(A,B,E)$: elements of $\mathcal{S}$ form the partition $A$, elements of $U$ are the partition $B$ and there is an edge between a vertex $v$ in partition $A$ and a vertex $u$ in partition $B$ if $v$ appears in the set $u$. The network is the graph $H$ and the problem is to find a dominating set/set cover, of a minimum size from partition $A$. This problem has a deterministic $O(f)$-approximation \congest algorithm in $O(f \log \Delta)$ rounds, when the frequency of each element is bounded by $f$. We will employ the following known result. 
	
	\begin{theorem}[Theorem 1 of Even et al.~\cite{EvenGM18}]\label{thm:setcover}
		There is a deterministic distributed algorithm in the \congest model that
		computes a $f(1 +\epsilon)$ approximation of minimum set-cover, in $O(\frac{\log (f\Delta)}{\epsilon \log\log (f\Delta) })$ rounds, in any set-system of frequency $f$ and maximum set size $\Delta$, and for any $0 <\epsilon< 1$.
	\end{theorem}
	\section{Deterministic \congest Algorithm}
	As already explained in the introduction a similar idea to what we will explain in the following was described in the work of Lenzen and Wattenhofer, but they used Luby's algorithm as a subroutine to obtain the desired approximation guarantee, here we will use the set cover problem to provide a deterministic \congest algorithm.
	
	To approximate MDS in rooted trees we can follow the following simple greedy choice: every parent chooses itself as a dominator. This gives a constant factor approximation for MDS on trees. However, for graphs of bounded arboricity, when there are multiple forests, this can cause an issue: too many parents are chosen for a specific set of children.
	
	To resolve this issue we convert the problem to the bounded frequency set cover problem. In the following, we explain how to perform this transformation in the \congest model. 
	
	\subsection{Construction of Set Cover Instance in the \congest Model}\label{sec3}
	The universe $U$ is the set of all nodes. We construct a bipartite graph $H=(A,B,E)$ (as an instance of distributed set cover) as follows.
	
	Let $\mathcal{F}$ be a forest decomposition with at most $f$ forests. 
	For a node $u$ we construct a set $S_u=\{u\}\bigcup C(u)$, here $C(u)$ is the set of children of $u$ in the forests of $\mathcal{F}$. We say $u$ is the \emph{representative} of set $S_u$. Then, the partition $A$ consists of all the possible sets of form $S_v$ for $v\in V(G)$. The part $B$ is just $V(G)$. The edge set $E$ is defined as the natural way: if an element (in part $B$) is in a specific set (of part $A$), then there is an edge between them. 
	
	To simulate the graph $H$ in the \congest model, every node $v$ simulates $S_v$, its image as an element in part $B$, and all of the edges from the set $S_v$ to the corresponding elements (children) in part $B$ as well as all edges from its image in part $B$ to all its container sets (parents) in part $A$. Since the forest decomposition is given in advance, every node can construct the corresponding subgraph of $H$ without an extra communication. Later when we invoke the set cover algorithm, whenever we are dealing with the set $S_v$ or an element $v$ in part $B$ the node $v$ of $G$ will perform the computational tasks.
	
	\noindent Observe that the degree of nodes in part $B$ is at most $f+1$, thus the instance has bounded frequency for a constant $f$.
	\begin{lemma}\label{lem:parents}
		Every element of $B$ appears in at most $f+1$ elements of $A$.
	\end{lemma}
	\begin{proof}
		Since there are $f$ forests, a vertex $u\in B$ is connected only to its own set $S_u$ or the corresponding sets of its parents in part $A$, so it has at most $f+1$ edges. Consequently it appears in at most $f+1$ sets.
	\end{proof}
	
	\subsection{Algorithm and Proof of its Correctness}
	
	We may assume that the input graph $G$ has arboricity at most $\alpha$. Then the algorithm is as follows.\clearpage
	
	\begin{algorithm}[h]
		\caption{\congest approximation of MDS on bounded arboricity graphs.\\ Input is a graph $G$ with arboricity at most $\alpha$.}
		
		\begin{algorithmic}[1]
			
			\STATE Construct a $(2+1)\alpha$-forest decomposition $\mathcal{F}$ by algorithm in~\cite{BarenboimE10}.
			
			\STATE Construct $H$ from $\mathcal{F}$ as explained in Section~\ref{sec3}.
			
			\STATE Compute $S$, an $O(\alpha)$-approximate set cover of $H$ by Theorem~\ref{thm:setcover}.
			\STATE Output the representative nodes of $S$.
		\end{algorithmic}\label{alg:mds}
	\end{algorithm}

	For the number of rounds we have the following lemma. 
	\begin{lemma}\label{lem:runtime}
		The Algorithm~\ref{alg:mds} runs in $O(\log n)$ rounds of the \congest model.
	\end{lemma}
\begin{proof}
	Since every step of the algorithm can be performed in the \congest model by messages of size $O(\log n)$, the algorithm as a whole is a \congest algorithm. For the running time, the first line requires $O(\log n)$ rounds to compute the forest decomposition. Once the forest decomposition is calculated, we have the second line for free, i.e.\ it can be computed without any new communication. For the third line, the running time depends on the choice of $\epsilon$-- preciseness of the approximation for set cover. Hence, for constant values of $\epsilon$, it takes $O(\frac{\log (\alpha\Delta)}{\log\log (\alpha\Delta) })$ rounds to perform the third line. Since $\Delta \le n$, the dominating factor in the round complexity is the first line, hence the algorithm performs in $O(\log n)$ rounds as claimed. 
\end{proof}
	
	Next we analyze the correctness and the approximation guarantee of algorithm. 
	
	\begin{lemma}\label{lem:correctness}
		The Algorithm~\ref{alg:mds} outputs a dominating set of $G$, which is an $O(\alpha^2)$-approximation for the MDS problem.
	\end{lemma}
	
	\begin{proof}
		Since every set cover of $H$ corresponds to a dominating set in $G$, the algorithm returns a valid dominating set. It remains to prove the correctness of approximation guarantee.
		
		Let $M$ be an MDS of $G$; we construct a dominating set $D$ that represents a set cover of $H$. By keeping the size of $D$ within $O(\alpha|M|)$ we ensure that every $O(\alpha)$-approximate set cover in $H$ is an $O(\alpha^2)$-approximation for MDS in $G$.
		
		Let $u\in M$ then $u$ dominates subset $S'_u\in V(G)$. Since it dominates all nodes in $S'_u$, it has edges to all of them. W.r.t.\ $\mathcal{F}$ these edges are either parent to child edges ($u$ is the parent) or child to parent edges ($u$ is a child). For the first case, $u$ has all its children in its corresponding bag $S_u$ in the set cover instance, it remains to add the latter nodes to our dominating set $D$ so that the union of their corresponding sets is the entire graph. Define $D\coloneqq M \bigcup_{u\in M}P(u)$, where $P(u)$ is the set of parents of $u$ w.r.t.\ forests of $\mathcal{F}$.
		Hence, we have $\bigcup_{v\in D}S_v = V(G) = U$. Thus $D$ is representing a set cover of $H$.
		
		Since by construction every node has at most $3\alpha$ parents, the size of $D$ is at most $(3\alpha+1)|M|$. 	 
		Therefore, as $D$ is representing a set cover of $H$, the optimal set cover has at most $|D| \le (3\alpha+1)|M|$ sets. Consequently, the $O(\alpha)$-approximation of set cover in line~3 of Algorithm~\ref{alg:mds} has size at most $O(\alpha (3\alpha+1)|M|)$, thus it corresponds to an $O(\alpha^2)$-approximation of MDS for $G$. 
	\end{proof}
	
	Our main theorem is a follow up of the previous lemmas.
	
	\begin{apthm}{thm:algorithm}
		There is a \congest algorithm that runs in $O(\log n)$ rounds and computes an $O(\alpha^2)$-approximation of MDS in graphs of arboricity at most $\alpha$.
	\end{apthm}
	\begin{proof}
		Algorithm~\ref{alg:mds} as proved in Lemma~\ref{lem:correctness} provides an $O(\alpha^2)$-approximation of MDS in a deterministic fashion. On the other hand by Lemma~\ref{lem:runtime}, it terminates in $O(\log n)$ rounds.
	\end{proof}
	
	To speed up the algorithm, we can use more forests to stop forest decomposition computation earlier, but then the approximation guarantee increases. For instance we can find a forest decomposition with $\sqrt{\log n}$ forests in $O\big(\frac{\log n}{\log \log n}\big)$ rounds to obtain the following. Here we only have to replace $\alpha$ with $\sqrt{\log n}$ in the Algorithm~\ref{alg:mds}. The following breaks the logarithmic round complexity barrier by sacrificing the approximation guarantee.
	
	\begin{corollary}
		There is a \congest algorithm that runs in $O\big(\frac{\log n}{\log \log n}\big)$ rounds and computes a $O(\log n)$-approximation of MDS in graphs of bounded arboricity.
	\end{corollary}
	
	%
	\textbf{Can the approximation factor be improved to $\alpha^2 +O(\alpha)$ or better?} We obtained almost $3\alpha^2+O(\alpha)$-approximation guarantee and our result was dependent on the existing algorithms for set cover and forest decomposition. We set $\epsilon=1$ in the algorithm for computation of the forest decomposition thus we got this bound; clearly one can use a smaller value for epsilon to achieve a better approximation guarantee (in the cost of increasing number of rounds proportional to $1/\epsilon$). 
	
	However, by this method, i.e.\ transferring the problem to set cover instance, it seems that it is impossible to provide a significantly better approximation guarantee. By this approach, unless there are logarithmic round algorithms to compute a forest decomposition with $\alpha + O(1)$ forests and an $(\alpha+O(1))$-approximation for the set cover problem, we cannot get a $\alpha^2+O(\alpha)$-approximation; we are not aware of existence of such algorithms. It is worth to mention that although the statement of the Theorem~1 in~\cite{ds-arbor} might be misinterpreted as a $(\alpha^2+O(\alpha))$-approximation guarantee; their distributed algorithm with existing tools has essentially a same approximation guarantee as ours.
	
	\section{Conclusion and Further Research Directions}
	We provided a deterministic \congest algorithm to constant factor approximate MDS on graphs of bounded arboricity and closed the gap between randomized and deterministic algorithms on these graphs. However, it is not clear what is the gap between possibilities and impossibilities for the MDS problem in graphs of bounded arboricity. Bad news is that there is no explicit lowerbound for these graphs. All we know about lowerbounds is either restricted to planar graphs~\cite{HilkeLS13,logstar} or is for general graphs~\cite{Kuhn:2016}, nothing in-between is known.

	Due to the algorithmic barrier involved with construction of forest decomposition, it might be the case that any radical improvement in the number of rounds for MDS could cause a fundamental improvement on the existing distributed tools for bounded arboricity graphs. However, there seems to be more accessible research directions. Maybe the first step is to improve the approximation factor: the best known sequential algorithm has $O(\alpha)$-approximation guarantee~\cite{BansalU17}\footnote{In their paper~\cite{ds-arbor} it is mentioned that the approximation guarantee of their sequential algorithm is $O(\alpha)$, however, after a personal communication with Christoph Lenzen it seems that the correct approximation guarantee for their sequential algorithm is $O(\alpha^2)$. Thus, to the best of our understanding the first $O(\alpha)$-approximation of MDS on graphs of arboricity $\alpha$ appeared in~\cite{BansalU17}}. Any $o(\alpha^2)$-approximation distributed algorithm is making the gap between sequential and distributed algorithms smaller. 
	
	\medskip 
	\paragraph{}
	
	\textbf{Acknowledgment: }We are grateful for wonderful comments of Yannic Maus and Sebastian Siebertz for the earlier draft version of this paper, specially for pointing out a flaw in the referencing and consequently the round complexity of the algorithm.
	\clearpage
	
		\bibliographystyle{abbrv}
	
	\bibliography{references}
	
\end{document}